\documentclass[preprint,pre,showpacs]{revtex4}

\usepackage{longtable}
\usepackage{graphics}
\usepackage{amsmath}
\usepackage{amssymb}
\usepackage{amsfonts}
\usepackage{bm}
\usepackage{rotate}

\usepackage{epsfig}
\usepackage[american]{babel}
\usepackage[dvipdfm]{color}
\usepackage[latin2]{inputenc}
\usepackage[american]{babel}
\usepackage{color}
\usepackage{graphicx}

\begin{document}
\title{Equilibriumlike extension of the invaded cluster algorithm}

\author{I. Balog}
\email{balog@ifs.hr}
\author{K. Uzelac}
\email{katarina@ifs.hr}

\affiliation{Institute of Physics, P.O.Box 304, Bijeni\v{c}ka cesta 46, HR-10001 Zagreb, Croatia}

\begin{abstract}
We propose an extension of the nonequilibrium invaded cluster (IC) algorithm, which reestablishes a correct scaling of fluctuations at criticality and also self-adjusts to the critical temperature. We show that by introducing a single constraint to the intrinsic quantity of the IC algorithm the temperature becomes well defined and the sampling of the equilibrium ensemble is regained. The procedure is applied to the Potts model in two and three dimensions.

\end{abstract}



\maketitle

\section{Introduction\label{secintro}}

 Monte Carlo (MC) studies of phase transitions have given rise to several cluster algorithms such as the Swendsen-Wang (SW) \cite{SwendsenWang} or Wolff algorithm \cite{Wolff}, which brought an important improvement to simulations by reducing the critical slowing down. More recently another cluster algorithm was proposed by Machta \textit{et al.} \cite{MachtaChayes}, inspired by the invasion percolation with an additional advantage that it self-regulates to the critical point and does not require prior knowledge of the critical temperature. It was applied to various classical models of phase transitions \cite{MachtaChayes,JGMC97}, and extended to more complex cases, from frustration \cite{FCC98} to tricritical points \cite{BU07}. The price to be paid is that the configurations generated by this self-adjusting nonequilibrium process are not distributed according to the canonical ensemble \cite{MachtaChayes,Moriarty}. Thus, although the IC algorithm can be used to analyze a certain number of properties at criticality, those which are describing fluctuations remain out of its reach. This question was analyzed from different aspects \cite{Moriarty,GLL05} and an alternative self-adjusting algorithm based on a modification of the Swendsen-Wang algorithm was proposed \cite{TomitaOkabe}.

 By following a different approach, we propose an equilibriumlike invaded cluster (EIC) algorithm obtained by constraining temperature uncertainty characteristic to the invaded cluster (IC) algorithm into limits compatible with the equilibrium distribution. We show that by applying this single constraint to the IC algorithm, correct scaling properties of thermodynamic observables are reproduced, while the algorithm still gives the critical temperature as an output. The method is demonstrated on the example of the Potts model in two and three dimensions.

 We consider the Potts model \cite{Potts} defined by the Hamiltonian
\begin{equation}
\label{dil_Potts}
H=-J\sum_{<i,j>}\big(\delta_{s_i,s_j}-1\big),
\end{equation}
where $s_i$ denotes the $q$-state Potts variable at the lattice site $i$ and the summation runs over the nearest neighbors only. The cluster algorithms including the IC algorithm are based on the Fortuin and Kasteleyn (FK) \cite{FortuinKasteleyn} expansion, which shows that the partition function of the model (\ref{dil_Potts}) is equivalent to the one of the random-cluster (RC) model, which can be written as
\begin{equation}
\label{rand_cluster}
Z=\sum_{\gamma\in \Gamma}p^{b(\gamma)}(1-p)^{B-b(\gamma)}q^{c(\gamma)},
\end{equation}
and understood as a generalized bond percolation, where
\begin{equation}
\label{p}
p=1-e^{-\beta J}
\end{equation}
is the bond probability ($\beta$ is the Boltzmann factor). The summation in Eq. (\ref{rand_cluster}) runs over the set of all the graphs on the lattice $\Gamma$, while each graph $\gamma$ represents one possible bond configuration. $B$ is the number of the lattice edges, $b(\gamma)$ denotes the number of bonds,  and $c(\gamma)$ is the number of connected components (FK clusters) in the graph $\gamma$.
\section{Algorithm\label{secalg}}

 In standard cluster algorithms such as the SW or Wolff algorithm, FK clusters are formed on a lattice by adding bonds between neighbors in the same state with a temperature-dependent probability $p$ (\ref{p}). The configuration is then randomized by flipping all the clusters to different states and the procedure is repeated from the beginning. In the IC algorithm by Machta \textit{et al.} the cluster formation phase is different since the probability $p$ is not given in advance. Bonds are placed randomly between neighbors in the same state, until one of the clusters percolates. This is considered as a signature of the critical point in the system of finite size. The adding of bonds then stops, clusters are randomized and ready for the next iteration. The output that is recorded is the ratio 
\begin{equation}
\label{p_gamma}
p_\gamma=\frac{b(\gamma)}{e(\gamma)},
\end{equation}
\noindent where $b(\gamma)$ is the number of added bonds  and $e(\gamma)$ the number of neighboring pairs in the same state for a given graph $\gamma$ on a lattice. The average $\overline{p}_\gamma$ is identified with the quantity $p$, yielding a critical temperature estimate. The procedure is self-adjusting to the critical temperature imposed by the stopping condition. Namely, if during one step the excessive number of bonds is added, which would statistically contribute to a much lower temperature, in the next step the clusters will percolate easier due to large number of satisfied neighbors created, which will statistically contribute to temperatures much higher than critical. The problem is that the resulting oscillations in $p_\gamma$ remain too large during the entire process. As a result it does not sample the canonical ensemble, but its proper {\it IC ensemble}, giving much broader distribution. Since in the thermodynamic limit $\sqrt{var(E)}/E \rightarrow 0$ still holds, the IC algorithm is assumed to become equivalent in that limit to the canonical ensemble and to produce the same average values, including the correct $T_c$. However, the convergence with size is more difficult to control. Also, the variances in energy or magnetization do not describe the equilibrium fluctuations, but rather the specific IC dynamics.

It is important here to notice that the variable $p$ of Eq. (\ref{rand_cluster}) is an {\it a priori} bond probability and consequently $p_\gamma$ should be distributed according to Gaussian distribution with $\overline{p}_{\gamma\;RC}=p$ and the width $\sqrt{var(p_\gamma)|_{RC}}\propto L^{-d/2}$. In the equilibrium cluster algorithms $p_\gamma$ appears as a random input variable which obviously is distributed according to normal distribution. As already mentioned, the distribution of $p_\gamma$ generated as an output of the IC algorithm is far broader than $L^{-d/2}$ \cite{MachtaChayes,Moriarty}. 

We propose a simple restriction to the IC algorithm, which sets the allowed range of $p_\gamma$ to be proportional to $L^{-d/2}$ and strictly less than $\sqrt{var(p_\gamma)|_{RC}}$. The procedure is straightforward and works as follows. The iterations are grouped in intervals of $N_a$ MC steps.

(a) During the first interval of $N_a$ MC steps we follow the original IC rule. The average value $\overline{p}_{\gamma,0}$ over the first $N_a$ iterations is found, to be used as an approximation for the next $N_a$ steps.

(b) In the next interval of $N_a$ MC steps the stopping rule is modified by requiring the system to percolate but with the constraint $\overline{p}_{\gamma,0}-v<p_{\gamma}< \overline{p}_{\gamma,0}+v$, where $v$ is the parameter set to be proportional to $ L^{-d/2}$. If the system percolates before the lower bound of $p_{\gamma}$  is reached, the bonds are still added until the lower bound is attained. If the upper bound of $p_\gamma$ is reached before percolation, the process is stopped without requiring the system to percolate. At the end of the interval, the new average over $p_\gamma$, $\overline{p}_{\gamma,1}$ is calculated.

(c) In every consecutive interval $i$ of $N_a$ steps the bounds of $p_\gamma$ are set by $\overline{p}_{\gamma,{i-1}}-v <p_\gamma < \overline{p}_{\gamma,{i-1}}+v$ and the same stopping rule as in (b) applied.

The first few intervals of $N_a$ steps are discarded and the configurations are recorded after the steady state has been reached. 

We remark that the width of the resulting distribution of $p_\gamma$ has two contributions: (a) from the fluctuations of $p_\gamma$ within a group of $N_a$ MC steps proportional to $v\propto L^{-d/2}$; (b) from the fluctuations of the mean value $\overline{p}_{\gamma,i}$ that correspond to the actual temperature fluctuations, producing the width of order $L^{-1/\nu}<L^{-d/2}$ if the heat capacity exponent $\alpha>0$, or of order $L^{-d/2}$ if $\alpha<0$. Even when it decays as $L^{-d/2}$, it is much smaller than the contribution from (a) because $\overline{p}_{\gamma,i}$ is a result from averaging over $N_a$ MC steps. The value of $N_a$ is required to be large enough to allow the uncertainty of values of $\overline{p}_{\gamma,i}$ to be less than the variation of $\overline{p}_{\gamma,i}$ between the groups $i$. The $\sqrt{var(p_\gamma)}$ of EIC algorithm thus can be regulated by changing $v$ and is proportional to $L^{-d/2}$. 

 Consequently, the width of the distribution of $p_{\gamma}$ allowed during the entire process of iterations is within the same limits as in the conventional cluster algorithms, and we can consider that the temperature during the whole process is equally well defined.

The question of the ensemble that EIC algorithm generates is less trivial and we do not attempt here any detailed study of the exact connection between the EIC and the canonical ensemble. We only point out that there is a nonzero probability of generating any spin configuration with any value of $p_\gamma$, so the ergodicity of the procedure is fulfilled. The numerical evidence will be given below that the EIC algorithm samples the canonical ensemble well. 
\section{Results\label{secres}}

The EIC algorithm was applied to the cases $q=2,3$ and $q=2$ of the Potts model in two and three dimensions, respectively. In two dimensions ($2D$) exact results are known both for critical temperature and for the exponents, while in $3D$ very accurate estimations are available \cite{Pelissetto}. The simulations were performed on lattices with periodic boundary conditions, and moderate sizes up to $L=226$ and $L=64$ in 2D and 3D, respectively. The statistics varied from $10^6$ iterations for smaller to $2\cdot 10^5$ iterations for the largest lattice sizes. Free parameters of the algorithm were chosen to be $v={L^{-d/2}}/{10}$ and interval $N_a=100$ Monte Carlo steps (MCS). Percolation was established by the topological rule, i.e. by the condition that the infinite cluster wraps around the lattice.
%
\begin{figure}[!hbt]
\includegraphics[scale=0.9]{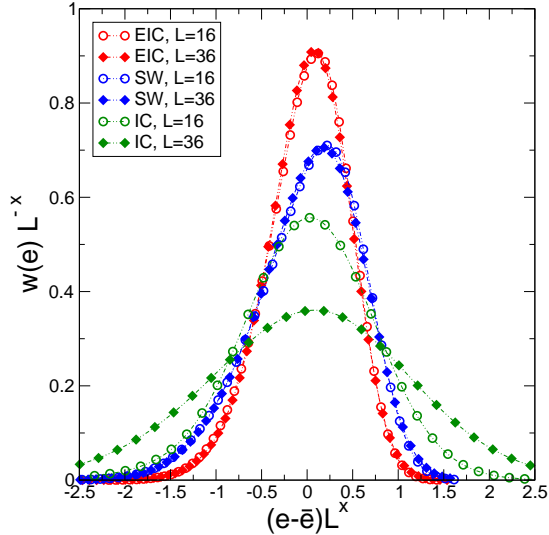}
\caption{(Color online) Energy histograms $w(e)$ of the EIC, SW and IC algorithm for the 3D Ising model and sizes $L=16$ and $L=36$ rescaled with the same exponent $x \approx 1.34$. $\overline{e}$ denotes the averaged energy of each individual curve. Histograms of SW algorithm were taken at $p_c(L)|_{EIC}$.}
\label{fig1}
\end{figure}
%
%

We include for comparison the results obtained by the SW algorithm with the same lattice sizes and using the same statistics performed at two different temperatures: (a) at infinite system critical temperature; i.e., for $p=p_c(L\rightarrow\infty)$; (b) at  L-dependent critical temperatures; i.e. for $p_c(L)$ approximated by previously calculated  $p_c(L)|_{EIC}$. Such a choice  is justified because the difference between $p_c(L)|_{EIC}$ and the position of e.g. the susceptibility maximum obtained by the SW algorithm for the lattice size $L$ is much smaller than its width.

In Fig. \ref{fig1} are compared the energy histograms, generated by EIC, SW and IC algorithms for the case $q=2$, $D=3$. All three sets were taken for sizes $L=16$ and $L=36$ and rescaled with the same exponent $x$, set to produce the collapsing fit of the SW data. The distribution generated by EIC algorithm is the narrowest one, due to the choice of the arbitrary constant factor in the parameter $v$. More important is that the width of the EIC histogram scales with the same exponent as the one of the canonical ensemble obtained by SW algorithm, so that the fluctuations in both ensembles follow the same law. Also, by tuning the constant factor in $v$, very good overlap can be established between the EIC and SW histograms of all sizes. On the other hand, the width of the IC ensemble scales differently, consistent with the fact that the corresponding fluctuations follow the specific IC dynamics, and do not describe the equilibrium energy fluctuations at criticality.

Let us present now the obtained results for critical point and critical exponents summarized in Table I.

\begin{figure}[!ht]
\includegraphics[scale=0.9]{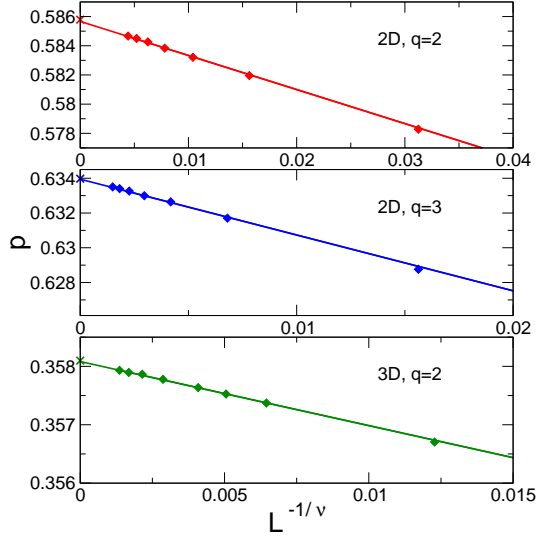}
\caption{(Color online) Data of $p_c(L)$ for cases $D=2$, $q=2$, $3$ and $D=3$, $q=2$. Lines represent linear fits, using exact or best known results for $\frac{1}{\nu}$.}
\label{fig2}
\end{figure}

First we examine the convergence of $p_c(L)$ with size. In Fig. \ref{fig2} are presented data for $p_c(L)$ vs $L^{-1/\nu}$, where the exact (or best known) value for $\nu$ was used. The obtained linear fits are very clean in comparison to the extrapolations within the IC algorithm (cf., e.g., Fig. 1 of the first Ref. in \cite{MachtaChayes}), where the use of medians was necessary to reduce the additional size effects. The extrapolations to the limit $L \rightarrow \infty$ are compared in Table I to the known results for $p_c$ in $2D$ (calculated from the analytic expression $p_c=1-(1+\sqrt{q})^{-1}$ \cite{Wu}) and in $3D$  \cite{Pelissetto} and show the precision up to four or five significant digits already for modest sizes used here.
%
%
\begin{table}[!h]
\begin{ruledtabular}
\caption{Results for critical exponents obtained by EIC (from $L=64$ to $226$ 
for 2D and from $L=16$ to $64$ for 3D) compared to those by SW algorithm 
obtained for the same lattice sizes  with the same statistics, and to the 
exact, or best approximate results.}
\centering{
\begin{tabular*}{8.2cm}{@{}l@{}c@{}c@{}c@{}c@{}c@{}}
 ~~ & $p_c$ & $y_\tau $ & $y_h$ & $\frac{\beta}{\nu}$ & $\frac{\gamma}{\nu}$ \\
\noalign{\smallskip}\hline
\colrule
  2D, $q=2$ \\
\hline
  EIC & $0.58575(2)$ & $0.98(1)$ & $1.868(6)$ & $0.132(5)$ & $1.775(7)$ \\
  SW\footnotemark[1] & - & - & $1.875(4)$ & $0.125(3)$ & $1.765(5)$ \\
  SW\footnotemark[2] & - & - & $1.876(5)$ & $0.127(4)$ & $1.750(5)$ \\
 exact \footnotemark[3]&$0.585786\ldots$ & $1$ & $\frac{15}{8}$&  $\frac{1}{8}$ & $\frac{7}{4}$ \\
\noalign{\smallskip}\hline
 2D, $q=3$ \\
\hline
 EIC & $0.63397(2)$ & $1.19(2)$& $1.861(6)$ & $0.140(5)$ & $1.745(7)$ \\
 SW\footnotemark[1]& -  & - & $1.86(5)$ & $0.136(5)$ & $1.750(8)$ \\
 SW\footnotemark[2]& -  & - & $1.88(4)$ & $0.120(4)$ & $1.740(5)$ \\
 exact\footnotemark[3]& $0.633974\ldots$ & $\frac{6}{5}$ & $\frac{28}{15}$ & $\frac{2}{15}$ & $\frac{26}{15}$ \\
\noalign{\smallskip}\hline
 3D, $q=2$  \\
\hline
 EIC &$0.35809(1)$ &  $1.590(3)$ & $2.481(3)$ & $0.520(2)$ & $1.987(4)$ \\
 SW\footnotemark[1]&- & - & $2.495(5)$ & $0.506(5)$ & $1.992(5)$ \\
 SW\footnotemark[2]&- & - & $2.490(5)$ & $0.510(4)$ & $1.989(5)$ \\
 \cite{Ferrenberg,Pelissetto}\footnotemark[4] & $0.358098(3)$ & $1.587(1)$& $2.482(1)$ & $0.5181(5)$ & $1.963(1)$ \\
\end{tabular*}
}\end{ruledtabular}
\footnotetext[1]{Simulations at $p_c(L\rightarrow\infty)$.}
\footnotetext[2]{Simulations at $p_c(L)|_{EIC}$.}
\footnotetext[3]{Exact values \cite{Wu}.}
\footnotetext[4]{Best known values for critical point \cite{Ferrenberg} and exponents \cite{Pelissetto}.}
\label{tab1}
\end{table}
%
The estimations of the temperature critical exponent, $1/\nu$ were obtained from the power law form $p_c(L)-p_c(\infty) \propto L^{-1/\nu}$ by taking the exact (best known) value for $p_c(\infty)$. 

The magnetic critical exponent was calculated in three different ways. We have calculated the fractal dimension of the percolation cluster $\overline{s}_{max}|_{p=p_c} \propto L^{y_h}$ and the anomalous dimension of the magnetization $\overline{m}|_{p=p_c}\propto L^{\beta/\nu}$ defined through the most populated of the q Potts states $a$, i.e.
\begin{equation}
\label{m}
m = \frac{q}{(q-1)\;L^d}\; max_{a} \left[ \sum_{i}\; \left(\delta_{s_i,a} - \frac{1}{q} \right) \right]. \\
\end{equation}
%
\begin{figure}[!htb]
\includegraphics[scale=0.9]{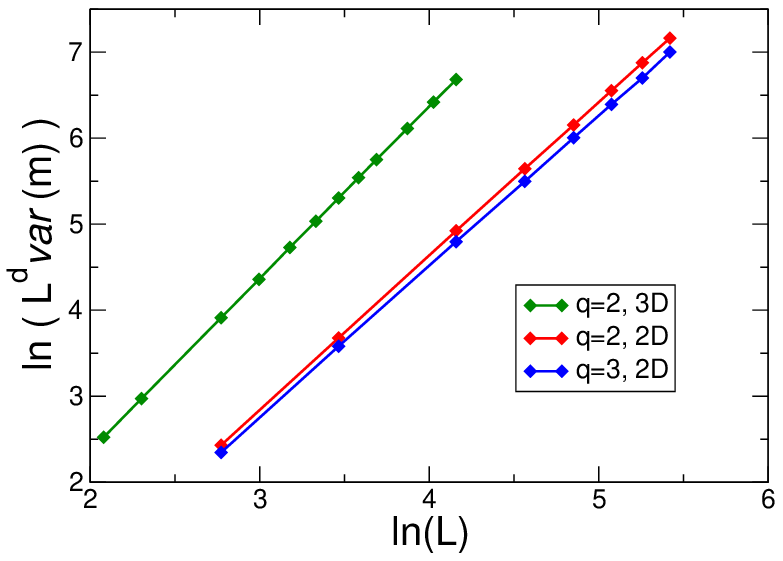}
\caption{(Color online) A ln-ln plot of the fluctuations of magnetization depending on the system size $L$.}
\label{fig3}
\end{figure}
%
We also examined the fluctuations of magnetization (see Fig \ref{fig3}) which in equilibrium are related to the susceptibility $\chi|_{p=p_c}\propto L^{\gamma/\nu}$ by the relation
\begin{equation}
\label{chi}
\chi = \; \frac{L^d}{k_B T} \; \left[\overline{m^2}-\overline{m}^2 \right],
\end{equation}
which assumes the validity of the fluctuation-dissipation theorem, not fulfilled for the standard IC algorithm.

EIC results for all the quantities presented in Table I agree with the exact, or best approximate results within a few percent. Although we omit here the analysis of convergence, visible improvement of results was observed with increasing L, so that discrepancies could be attributed to relatively modest sizes used. This is also supported by the fact that the simulations by SW algorithm with the same sizes produced very similar results. Compared to those of the standard IC algorithm \cite{MachtaChayes}, the fits of quantities derived from the mean values at criticality, such as $p_c$, $1/\nu$, and $\beta/\nu$ are significantly improved. Moreover, the correct exponent is regained for the fluctuations of the magnetization, and the susceptibility shows the correct scaling within the same error margins as within a SW algorithm. 
\section{Discussion and conclusion}

To summarize, we have proposed an extension of the IC algorithm which recovers the correct sampling of the equilibrium ensemble at criticality and still preserves self-regulation to the critical temperature. By imposing on the distribution of the variable $p_\gamma$ the width proportional to $L^{-d/2}$ (compatible with the Gaussian distribution), we reduced the uncertainty of the temperature variable to the one in the equilibrium algorithms and obtained the correct scaling of the fluctuations at criticality. Furthermore, our intervention does not slow down the IC algorithm, and the running times per MCS for the EIC algorithm remain the same. For example, a run of $10^5$ MCS for 2D Ising model on $L=64$ lattice requires approximately $500$ s on the AMD Opteron 240 processor (1.4 GHz). The procedure is illustrated on several cases of second-order phase transitions in the Potts model in two and three dimensions belonging to different universality classes.

In comparison with the self-regulating algorithm by Tomita and Okabe \cite{TomitaOkabe}, this approach is conceptually different: their approach is an extension of the SW algorithm which allows variation in $p$, which is directly related to the fluctuations of temperature, and the corresponding width is $\propto L^{-1/\nu}$. As far as the efficiency is compared, the authors of Ref. \cite{TomitaOkabe} argue their algorithm to be faster per individual iterations since it does not require multiple checking of percolation during a single iteration. On the other hand, in Ref. \cite{TomitaOkabe} not less than 10000 MCS preparation steps were required before the iterations could be recorded, while in the EIC approach not more than 2000 MCS were sufficient. Thus, it would be interesting to compare the autocorrelation times for the two methods.

In the future, more detailed study remains to be done of the ensemble, the leading convergence exponent, the autocorrelations, and the corresponding dynamic exponent. 

The improved convergence and possibility to calculate the correct scaling of fluctuations may be useful in a number of problems. It may have particular advantage in the cases with quenched disorder and lack of self-averaging, where calculations at the sample-dependent critical temperatures have to be performed, and where the standard IC algorithm was of limited efficiency \cite{BUprep}.

This work was supported by the Croatian Ministry of Science, Education and Sports through grant No. 035-0000000-3187.
\vfill

\end{document}